\documentclass[12pt]{article}
\usepackage[margin=0.5in]{geometry}
\usepackage[pdftex]{graphicx}
\usepackage[latin1]{inputenc}
\usepackage{amsmath}
\usepackage{amsfonts}
\usepackage{amssymb}
\usepackage{times}

{

\title{Swarming in disordered environments}
\author
{David Quint and Ajay Gopinathan\\
\\
\normalsize{Department of Physics, University of California Merced, Merced CA 95343 USA}\\
}

\date{}

\begin{document}

{\setlength{\parindent}{0cm}
\baselineskip24pt
\maketitle

\begin{abstract}
\noindent The emergence of collective motion, also known as flocking or swarming, in groups of moving individuals who orient themselves using only information from their neighbors is a very general phenomenon that is manifested at multiple spatial and temporal scales. Swarms that occur in natural environments typically have to contend with spatial disorder such as obstacles that hinder an individual's motion or communication with neighbors. We study swarming particles, with both aligning and repulsive interactions, on percolated networks where topological disorder is modeled by the random removal of lattice bonds. We find that an infinitesimal amount of disorder can completely suppress swarming for particles that utilize only alignment interactions suggesting that alignment alone is insufficient. The addition of repulsive forces between particles produces a critical phase transition from a collectively moving swarm to a disordered gas-like state. This novel phase transition is entirely driven by the amount of topological disorder in the particles environment and displays critical features that are similar to those of 2D percolation, while occurring at a value of disorder that is far from the percolation critical point. 
\end{abstract}

\section*{Introduction}
Collective motion of self propelled individuals is a well studied emergent phenomenon~\cite{vicsek3} that spans many different length and time scales from biopolymers on a bed of molecular motors~\cite{Bausch}, swimming bacteria~\cite{Zhang, Aparna}, birds and fish~\cite{Couzin1,Couzin2} to people ``moshing" at heavy metal concerts~\cite{Silverberg}. Within the literature that is aimed at studying collective motion in systems of self-propelled particles~\cite{vicsek1,vicsek2,vicsek4,vicsek5,orsogna,yates}, a main underlying assumption has been that the environment, where the particles exists, is continuous, isotopic and ordered. In the natural world there are many examples of disordered environments where collective motion can exist. Examples include bats which navigate natural caverns via echolocation, schools of fish that maneuver through dense kelp forests, microbial colonies that move about in heterogeneous soil~\cite{Gilligan}, crowds of people that are evacuating a building~\cite{Helbing1} and traffic flow in major cities~\cite{Helbing2}. Given that natural environments can be intrinsically disordered, it is interesting to consider how self propelled individuals maintain an organized state of collective motion without knowledge of a  global ``road map". What are the necessary physical mechanisms that facilitate the collective movement of particles across a intrinsically disordered network? What is the role of "thermal" noise in a swarm that moves through a disordered environment? In this manuscript we provide insight into these questions for the first time, by studying a two dimensional system in which we represent the disordered environment as a percolated lattice that our self-propelled particles inhabit. 

\section*{Model $\&$ Simulation}
To study swarming behavior in the presence of topological disorder we implement a Monte Carlo lattice gas model~\cite{vicsek2,Frey} that consists of $N_{p}$ interacting mobile particles that occupy a $2d$ periodic triangular lattice with $L^2\equiv N$ lattice sites. The particle density is defined as $\rho=N_{p}/N$ and in this model it can, in general, be larger than the density of lattice sites. The dynamics of our model [see supplementary information for more details] are such that, at any given time step, particle $k$ moves along any of the six lattice bonds (with direction unit vectors $\mathbf{u}_i$) with a velocity $\mathbf{v}^{(k)}$ whose magnitude is a constant defined to be unity [bond length/time step] (Fig.~\ref{fig: figure1}A). Particle moves are controlled via a standard Monte Carlo procedure with Boltzmann weights determined by two interaction energies, between nearest neighbor (n.n) particles, in our model. The first is an alignment interaction energy, $E^a_{i}$ (Eqn.~\ref{eq:alignAVG}), that makes it favorable for particles to orient their velocity vectors along the direction of the average velocity of their n.n. The magnitude of alignment is controlled by the parameter $\alpha$, which is set to unity and defines the relevant energy scale. The second is a mutual repulsive interaction energy, $E^a_{i}$ (Eqn.~\ref{eq:repuls}), that is proportional to the difference in the local density $n(i)-n(0)$ seen by a particle, where $n(i)$ is the number of particles at a n.n. site along a lattice direction $\mathbf{u}_i$ and $n(0)$ is the number of particles at the particle's current site. The magnitude of the repulsive interaction is controlled by the parameter $\epsilon$ and the local repulsive energy enhances the probability of moving in directions with the greatest drop in density. We introduce topological disorder in our lattice system by varying $p$, the probability that a bond exists between two lattice sites as in usual bond percolation theory~\cite{Havlin}. The quantity, $1-p$, therefore  represents the environmental disorder fraction.  Finally, the parameter $T$, which enters via the Boltzmann weighting for the Monte Carlo procedure effectively controls the magnitude of \textit{thermal noise} in our system. We quench the disorder in the lattice during each realization of our simulation, allowing us to study the effect of topological disorder and thermal noise independently and their effect on the \textit{fidelity} of local information that particles use to navigate the disordered environment. 

\begin{equation}
E^a_{i}= -\alpha\mathbf{u}_i\cdot\sum_k^{n(j)}\mathbf{v}_k \label{eq:alignAVG}
\end{equation}

\begin{equation}
E^{r}_{i} = \epsilon (n(i)-n(0)) \label{eq:repuls}
\end{equation}

\begin{figure}[h]
\center \includegraphics[clip=true, trim=50 0 90 50,scale=0.45]{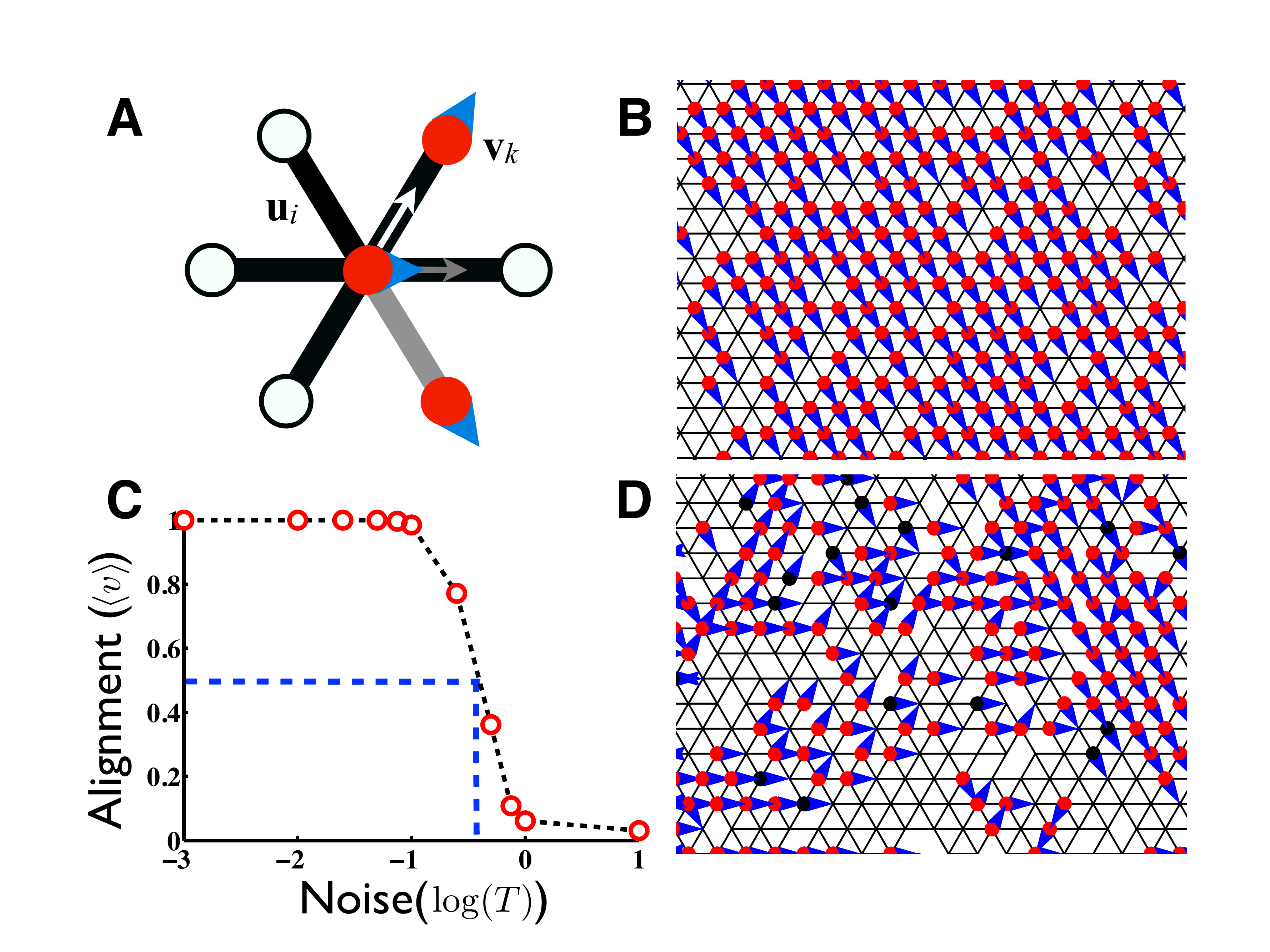} 
\caption{(\textbf{A}) Diagram of a lattice unit cell with particles (red) occupying lattice sites. Solid black bars are occupied lattice bonds and grey bars are deleted bonds. The blue triangles represent current particle velocity directions. White/Grey arrows indicate some potential lattice directions along which the center particle can move at the next time step. In the absence of any repulsion, if the deleted bond (grey bond) were still present, the updated direction of the center particle would most probably be along the grey arrow, since that is the local average velocity direction. However, since the bond is not present, the most probable direction is along the white arrow. Note that the absence of the bond not only disallows motion in that direction but also inhibits transfer of information. (see supplementary information)(\textbf{B})  A snapshot of a typical simulation showing finite sized groups of particles which are collectively moving in a disorder free ($1-p=0$) lattice. (\textbf{C}) Order-disorder transition driven by the thermal noise in a system of particles ($\rho=1.0$ and $1-p=0$) without repulsive interactions for a system size of $N=1024$. The blue dash line indicates the location of the critical transition temperature, $Tc \simeq 0.25$, which was determined numerically. (\textbf{D}) A snapshot of a simulation of a system with disorder ($1-p=0.05$), where missing bonds are not drawn. Black filled circles are particles that are temporarily stuck at a lattice bond defect.} 
\label{fig: figure1}
\end{figure}
 
\section*{Results}
In general, the existence of intrinsic noise in a system of mobile interacting particles can drive the departure from a collectively moving ordered swarm state to a completely disordered gas-like state state~\cite{Aldana}. In our system, we consider a different source of disorder, environmental topological disorder, which is controlled by the parameter $p$, in addition to thermal noise. Focusing first on the thermal noise dependence of our model without the introduction of any environmental disorder ($p=1$) (Fig.~\ref{fig: figure1}B), we find that there exists a order-disorder phase transition near a critical magnitude of thermal noise $T_c$ Fig.~\ref{fig: figure1}C. This transition is reflected in the sharp drop of the global alignment order parameter $\langle v \rangle$ (Eq.~\ref{eqn:orderP}) at $T_c$. Here $\langle v \rangle$ is the mean velocity of the $N_p$ particles in our system and can range from unity for a perfectly ordered swarm moving collectively along a single direction to zero when particle motions are completely uncorrelated.

\begin{equation}
\langle v \rangle = \frac{1}{N_{p}} \bigg\vert \sum_i^{N_{p}} \mathbf{v}_i\bigg\vert , \hspace*{5pt} \text{where} \hspace*{5pt}  \vert \mathbf{v}_i\vert=1
\label{eqn:orderP}
\end{equation}

The phase transition associated with this order parameter is in good agreement with previous studies of the classic Vicsek model~\cite{vicsek1,vicsek2}. 
To understand the effects of topological disorder on the formation of a collective swarm we fix the magnitude of thermal noise below the critical value, $T<T_c$. In doing so we hold the system in a regime where swarming would naturally occur without environmental disorder. First we consider the case where particles only interact via a local alignment interaction (Eqn.~\ref{eq:alignAVG}). In Fig.~\ref{fig: figure2}A we see the effect of environmental disorder on the formation of a collective swarm as characterized by the order parameter Eqn.~\ref{eqn:orderP}, averaged over multiple disordered lattice realizations, for a system without repulsive interactions. At low densities of particles ($\rho \le 0.2$), consistent with earlier studies ~\cite{vicsek2}, we find that there exists no collectively moving swarm ($\langle v \rangle<1$) that involves all particles in our system even for a perfectly ordered lattice. At higher densities, for low values of the disorder fraction, we find a completely ordered swarming state that exhibits a phase transition beyond some fraction of missing bonds ($1-p^\ast_0$) in our system. This order-disorder transition is completely determined by the amount of topological disorder, which is unrelated to the classic Vicsek model transition that is induced by thermal noise\cite{vicsek1,vicsek2,Aldana,Baglietto}. To understand if this transition uniquely defines a critical disorder fraction in the thermodynamic sense, we characterized the location of the critical disorder fraction ($1-p^\ast_0(L)$) for finite systems as a function of the system size.
Finite size scaling analysis revealed (Fig.~\ref{fig: figure2}B) that, in the large system size limit ($L\rightarrow \infty$), the existence of a critical phase transition that is governed by environmental disorder is fully suppressed, ($1-p^\ast_0(\infty)=0$). This result suggests that, in the absence of repulsive forces between particles, any amount of environmental disorder destroys the ability for a system of particles to collectively swarm. In contrast to this behavior, we now consider the effect of adding a mutually repulsive interaction between neighboring particles (Eqn.~\ref{eq:repuls}). In Fig.~\ref{fig: figure2}C, we see that the ability of particles to form a collectively moving swarm is significantly enhanced compared to Fig.~\ref{fig: figure2}A even for moderate values of the disorder fraction. At first glance it seems that the effect of adding a repulsive potential between particles has changed the location of where the order-disorder transition ($1-p^\ast(L)$) takes place along the disorder fraction axis for finite systems. Again, using finite size scaling, we found that there exists a true thermodynamic phase transition from an ordered swarm to a disordered one (Fig.~\ref{fig: figure2}B) that occurs at a finite value of the topological disorder fraction. For a specific value of the repulsive interaction, $\epsilon=10^{-1}$, and thermal noise $T=10^{-2}$, this turns out to be $1-p^\ast(\infty)=0.13(4)$. Furthermore, the location of the critical disorder fraction moves as the repulsive interaction is changed as shown in Fig.~\ref{fig: figure2}D. At a fixed value of thermal noise, the {\it location} of the critical phase transition is governed by ratio of the two interaction energies, $\epsilon/\alpha$ ($\alpha=1$), but the critical scaling behavior near the critical point may in fact be universal. The universality class is usually determined by the values of the critical exponents that govern the scaling of various physical quantities near the critical point.  We first extended our finite size scaling analysis to compute the critical exponents associated with this transition for a fixed value of $\epsilon=10^{-1}$. The correlation length $\xi$, which measures the length scale over which particle motion is correlated across the disordered lattice, diverges becoming comparable to the system size ($\xi\sim L \rightarrow (p^\ast(L)-p)\sim L^{-1/\nu}$) as we approach the critical disorder fraction transition. This allows us to compute the associated exponent $\nu$. From the fit to the scaling law in Fig.~\ref{fig: figure2}B, we find that $\nu=1.25(6)$.  We also measured the susceptibility of the order parameter to the lattice disorder fraction by measuring the peak value of the order parameter fluctuations as a function of system size~\cite{Baglietto}, $\chi_v = N\sigma_v^2(p^\ast)~\sim L^{\gamma/\nu}$. This provided an estimate for the susceptibility critical exponent, $\gamma=2.70(5)$ (Fig.~\ref{fig: supfig1}A). \par

 We also performed an independent measurement of the susceptibility exponent by examining how the fluctuations in the order parameter, $\chi = L^2\sigma_v^2(p)\sim(p^\ast(L)-p)^{-\gamma}$, scale with disorder fraction $1-p$ near the critical point for different values of the repulsion magnitude. We found that for over three decades of the repulsive interaction magnitude that the estimate for the critical exponent given by Fig.~\ref{fig: supfig1}A agreed very well with estimates given by the scaling near $p^\ast(L)$ (Fig.~\ref{fig: supfig1}B), suggesting universal behavior with respect to repulsive strength. Scaling of the the order parameter, $\langle v\rangle\sim (p^\ast(L)-p)^\beta$ with respect to $p$, revealed the scaling exponent $\beta$ (Fig.~\ref{fig: supfig1}C), which was also insensitive to the value of the repulsive interaction. Using the hyper scaling relation $2\beta +\gamma=d\nu$, we were able to infer the scaling exponent $\nu$ (Fig.~\ref{fig: supfig1}D) over the same repulsive interaction range and found a reasonable comparison with our estimates from finite system size scaling (Fig.~\ref{fig: figure2}C). To gain perspective on which universality class best describes swarming in disorder, we compared our exponents to those of two dimensional percolation~\cite{kesten} and the standard Vicsek model ~\cite{vicsek1,Baglietto}. We found that our system follows a universality class that is closer to that of percolation than the Vicsek type as shown in Fig.~\ref{fig: supfig1}B-D). It is interesting to note that this system is more akin to the percolation type universality class given that the critical disorder fraction ($1-p^\ast(\infty)=0.13(4)$) for a repulsive strength of $\epsilon=10^{-1}$ is vastly different than that of ordinary connectivity percolation ($1-p_c \sim 0.66)$) for a triangular lattice~\cite{Sykes}. This comparison suggests that swarming phenomena in such systems are extremely sensitive to the effects of ordinary percolation far from the percolation critical point while retaining the critical behavior of a percolation type system. It is also relevant to note that this disorder induced phase transition occurs in a self-propelled collective swarm rather than being a field driven transition as is common in condensed matter systems. \par 
 Our results show that a finite amount of repulsion enhances swarming, suggesting that there might be an optimal degree of repulsion depending on the system parameters. In Fig.~\ref{fig: figure3}A we examine the effect that repulsion has on the order parameter in the presence of different amounts of disorder. In general, for finite disorder, we found that repulsion has a non-monotonic effect on the ability for particles to form collectively moving swarms. When the magnitude of the repulsion is small, we find that there is no significant enhancement of swarming ability above what is seen for systems without repulsion. As the repulsion is increased, we find that there is a maximal enhancement of the order parameter, represented by the peaks in Fig.~\ref{fig: figure3}A. Beyond this maximum, we find that, for strong values of repulsion, there is a suppression of collective behavior, indicating the existence of an optimal repulsive interaction for a given disorder fraction.
 
To gain insight into this non-monotonic behavior, we look at the instantaneous disorder averaged mobility fraction, which measures the fraction of particles that are not temporarily stuck at a defect, for various values of $1-p$ (Fig~\ref{fig: figure3}B). Increasing the repulsion magnitude, we find that the mobility fraction also increases for a fixed disorder fraction. Visually we can confirm that particle motion is becoming less hindered by the existence of lattice defects as shown in the simulation snapshots Fig.~\ref{fig: supfig3}B and D when compared to Fig.~\ref{fig: supfig3}A and C. This result is also consistent with the increase in the critical disorder fraction as $\epsilon$ is increased at fixed thermal noise (Fig.~\ref{fig: figure2}D). As we approach the maximal repulsive magnitude, we find that the mobility fraction saturates at fixed disorder fraction (Fig.~\ref{fig: figure3}B) and the critical disorder fraction saturates as well (Fig.~\ref{fig: figure2}D). Increasing the repulsive interaction past the optimum value ($\epsilon\sim10^{0}$), we find that the mobility fraction remains maximal for $1-p\lesssim0.6$ (Fig.~\ref{fig: figure3}B), while the order parameter in this regime is greatly reduced (Fig.~\ref{fig: figure3}A) and there exists no collective state even for a disorder fraction equal to zero. It is interesting to point out for a large disorder fraction ($1-p\gtrsim 0.66$), there is a slight dip in the mobility fraction for repulsion  near $\epsilon \gtrsim 10^{-1}$. This is due to the geometric percolation transition for a triangular lattice, which has a critical disorder fraction of $1-p_c=1-2\sin(\pi/18)\simeq0.652 $~\cite{Sykes}. Near this transition, we find that the particles become trapped in local clusters in the lattice and have a slightly higher mobility fraction for $1-p>1-p_c$. At large values of repulsive strength ($\epsilon\sim10^1$) the mobile fraction is greatly reduced at even small values of the disorder fraction, because particles are forming a gas-like state that interacts with defects more frequently such that the mobility scales as the number of missing bonds in the lattice $\langle\mu\rangle\sim  1-p$ for $p\lesssim 1$ as opposed to $\langle\mu\rangle\sim1$ for $10^{-2}<\epsilon<10^{1}$. 

To summarize our findings, we present the phase diagram for fixed thermal noise  ($T < T_c$) defined by the magnitude of the order parameter (Eqn.~\ref{eqn:orderP}) in Fig.~\ref{fig: figure3}C. In general, we find that for both large values of the disorder fraction and the repulsion magnitude that swarming is suppressed. Moreover, for a fixed value of the disorder fraction we see that scanning through the repulsion magnitude (Fig.~\ref{fig: figure3}C vertical axis) from low to high takes the order parameter through a maximum, as in Fig.~\ref{fig: figure3}D. At larger values of thermal noise, we find that the broad enhanced region is washed away (Fig.~\ref{fig: supfig2}C). It is interesting to note that there is a weak enhancement effect from thermal noise, which can be seen in Fig.~\ref{fig: supfig2}A-C. We find that thermal noise ($T <  T_c$) allows the system to anneal in the presence of topological disorder, which pushes out the order-disorder transition (even in the absence of repulsive interactions(Fig.~\ref{fig: supfig3}B) to slightly larger values of the disorder fraction (Fig.\ref{fig: figure2}D). Here, due to thermal noise, particles are susceptible to random fluctuations in their ability to align with their neighbors thus allowing the system to behave more ergodically.

\section*{Conclusion}
Navigating an intrinsically disordered environment without \textit{a priori} knowledge of the topology of the environment possess a significant challenge for individual organisms. The ability to form a collective intelligence in the form of a swarm can greatly reduce the navigational complexity in a disordered environment. A swarm utilizes physical interactions between individuals as a means of a primitive collective perception in order to overcome environmental obstacles. The plasmodium of \textit{Phyarum polycephalum} when placed in a two dimensional maze uses a collective cellular oscillations to find the shortest path between food sources~\cite{Nakagaki1,Nakagaki2}. Microbial colonies navigate tiny interconnected channels formed by roots of plants~\cite{Gilligan}. How do these real systems deal with the disorder? How do they communicate local information about their environment to their neighbors? Topological disorder is also intrinsic in social networks, the internet and scientific citation networks~\cite{Albert}. One may ask the question what is the critical local degree of connectivity that will allow for the entire population to act holistically~\cite{Watts1,Watts2,Albert}? In robotics it is becoming more common to utilize robotic drones to explore dangerous environments in place of humans. What are the necessary interactions that groups of robotic drones must possess in order to perform tasks such as search and rescue in an environment where the local topology may not be known a priori~\cite{Lin}? Systems of self propelled particles using local nearest neighbor alignment individual particles can form a collectively moving swarm~\cite{vicsek1,vicsek2}. However, in the presence of environmental topological disorder alignment alone is insufficient for particles to form a swarm. In order for these systems to navigate disorder they must possess information not only from local neighbors directions of travel but also the local surrounding environment. Repulsive forces allow particles to communicate local topological features to their neighbors and restores the ability of the individuals to form a swarm that collectively navigates the intrinsically disorder environment. We have shown there exists a new type of dynamical phase transition driven by environmental disorder and that the ability for agents to collectively move in these disordered environments requires the presence of local repulsive forces between collectively moving particles. Collective motion in these disordered environments can be optimized by tuning the magnitude of the repulsive interaction for a given amount of disorder. In nature there may exist evolutionary pressures that select for better swarming ability within a group of individuals. It is interesting to speculate whether organisms that routinely deal with disordered environments have evolved mechanisms that effectively mediate repulsive interactions similar to the one we have studied here.

\begin{figure}[h]
\center \includegraphics[clip=true,trim=50 0 20 0, scale=0.5]{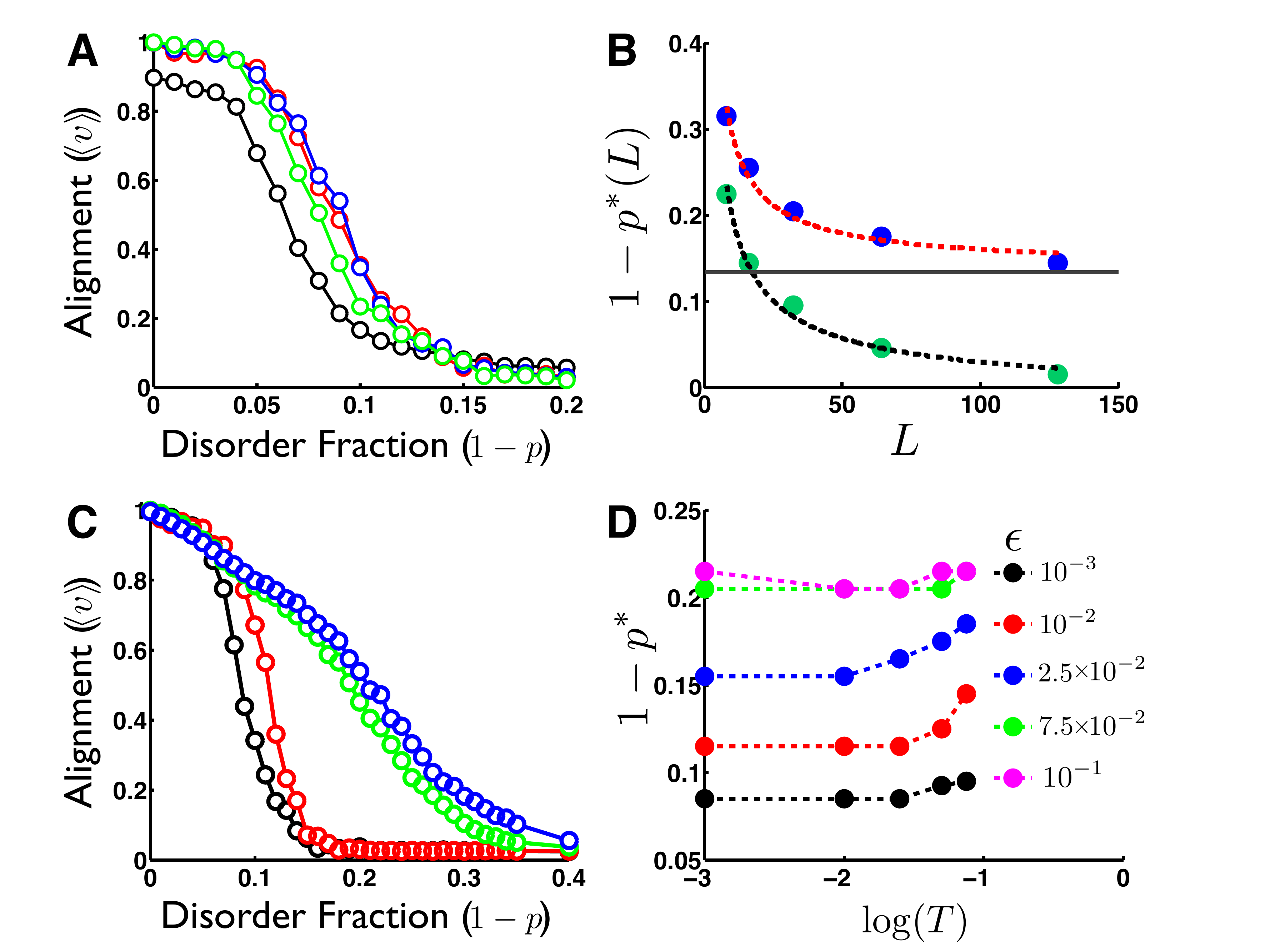} 
\caption{(\textbf{A}) Global alignment, in the absence of repulsion, as measured by Eqn.~\ref{eqn:orderP} as a function of the disorder fraction in the lattice for different densities [$\rho=0.2$ (black), $\rho=0.8$ (red), $\rho=1.0$ (blue), $\rho=2.0$ (green) ].  (\textbf{B}) Finite system size critical disorder fraction ($1-p^\ast(L)$) for systems without repulsion [$\epsilon=0$ (green filled circles)] and with repulsive interactions [$\epsilon>0$ (blue filled circles)]. Fitting to finite size scaling, $(p^\ast(L)-p^\ast(\infty))\sim L^{-1/\nu}$, predicted a critical disorder fraction for systems without repulsive interactions (black dashed line) that is zero and for systems with repulsion (red dashed line) is found to be $1-p^\ast(\infty)= 0.13(4)$(black solid line). The critical scaling exponent $\nu$ in the case for repulsive interactions is $1.25(6)$. (\textbf{C}) Global alignment for a system of particles that interact with both a local alignment and repulsive fields for $\rho=1.0$ and $\epsilon =[5.0*10^{-3}$ (black), $10^{-2}$ (red), $5.0*10^{-2}$ (blue), $10^{-1}$ (green)]. The location of the order-disorder transition is pushed out allowing particles to swarm in the presence of moderate environmental disorder. (\textbf{D}) The location of the critical disorder as a function of temperature at various values of the repulsive interaction strength (see inset). For low values of repulsion the critical disorder changes very little over temperature. At intermediate repulsion values we see there is an enhancement effect with increasing thermal noise. At repulsive strengths near maximal (Fig.~\ref{fig: figure1}D) we find little change of the critical disorder fractions over entire range in $T$. All data shown in (A), (C) and (D) is for a system size $N=1024$. All data in (A), (B) and (C) presented here was obtained at a thermal noise of $T=10^{-2}$}
\label{fig: figure2}
\end{figure}

\begin{figure}[h]
\center \includegraphics[clip=true, trim=0 0 0 0,scale=0.5]{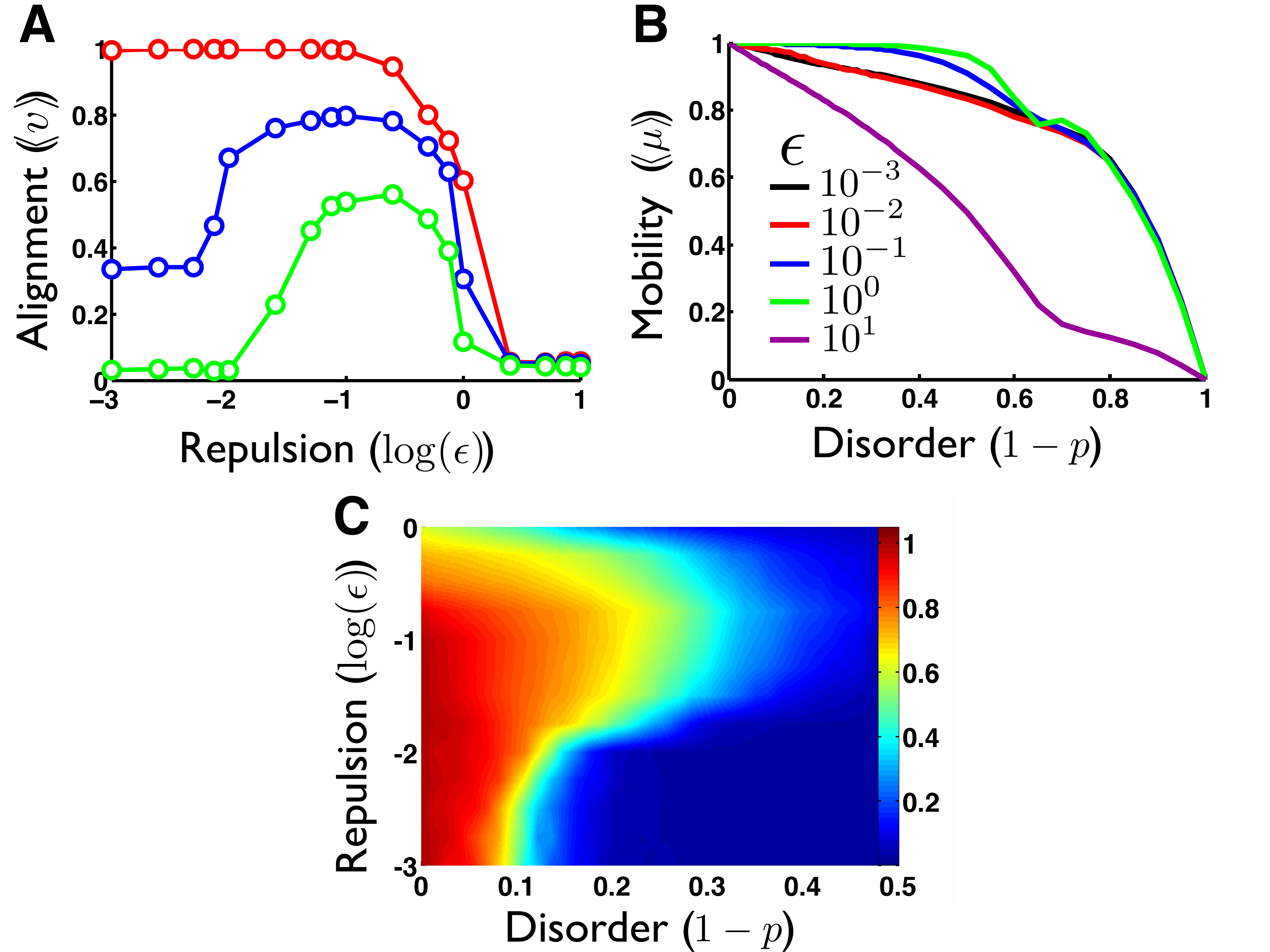} 
\caption{(\textbf{A}) Behavior of the order parameter for particular values of the disorder fraction [$p=1.0$ (red), $0.90$ (blue), $0.80$ (green)] clearly shows a non-monotonic dependence on the magnitude of the order parameter (log horizontal axis) with a central maximum for curves with $p<1$. (\textbf{B})The ensemble averaged instantaneous mobility of particles as function of environmental disorder. For smaller values of the disorder fraction, the effect of increasing the repulsive energy significantly increases particle mobility allowing particles to swarm. As the disorder fraction is increased, the magnitude of the mobility generally decreases as expected. Interestingly, for values of the disorder fraction near the connectivity bond percolation threshold ($1-p_c\simeq0.66$) we find a significant dip in the mobility of particles at large values of the repulsive energy parameter ($\epsilon\gtrsim 10^0$). This signifies that particles are becoming trapped in local disconnected clusters in the lattice. (\textbf{C}) Phase diagram defined by the magnitude (color bar) of the order parameter (Eqn.~\ref{eqn:orderP}) for the two parameters $\epsilon$ and $1-p$ at $T=10^{-3}$. The large peak centered near $\log(\epsilon)\sim -1$ shows the effect of repulsion has on the persistence of swarming the presence of disorder.}  
\label{fig: figure3}
\end{figure}

\section*{Supplementary Information}
\setcounter{equation}{0} 
\makeatletter 
\renewcommand{\thefigure}{S\@arabic\c@figure}  
\makeatother

\subsection*{Model-Monte Carlo}
In this section we detail our simulation methods. Lattice configurations were generated randomly for a given value of the disorder fraction $1-p$ and system size $N=L^2$. During any particular realization of the simulation, the bond topology was kept fixed. Particle updates were carried out at each time step by selecting a particle at random. The particle's direction of motion along one of the six lattice directions ($i=[1,6]$) was updated by computing the probability to move $P_i$ (Eqn.S~\ref{eq:probOr}) in a given direction, $\mathbf{u}_i$ using the local alignment field (Eqn.~\ref{eq:alignAVG}) and repulsive particle-particle number density (Eqn.~\ref{eq:repuls}). In general each of lattice sites has a number $n(j)\geq 0$ particles which maybe be overlapped and can point in any of the six lattice directions. Topological disorder is coded into the update probability, $P_i$ by the condition that when a bond exists between two neighboring lattice sites then $\eta_i=1$ and conversely when a bond is missing $\eta_i=0$. Trial moves were selected using a standard METROPOLIS Monte Carlo method where particle moves were accepted/rejected by comparing a randomly generated number $0\leq r\leq 1$ to a mapping of the move probabilities (Eqn.~\ref{eq:probOr}) to the interval [0,1]. Lattice directions that have a missing bond are assigned a move probability equal to zero. If a particle encounters a broken bond it will wait there until the next move update. The total number of particle updates to reach steady state are system size dependent and vary from $10^6$ for $N_{sw}=16^2$ to $10^8$ for $N_{sw}=128^2$. All averaged quantities are disorder averaged over different realizations of the lattice network at fixed values of $1-p$. 

\begin{equation}
P_i\equiv P(\mathbf{u}_i) = \frac{1}{\mathcal{Z}}\eta_{i}\prod_{j=1}^6  \exp\bigg[-\beta\bigg(-\alpha \eta_{j} \mathbf{u}_{i} \cdot \mathbf{f}(j)\bigg)\bigg]\times\exp(-\beta E^{r}_{i}),\label{eq:probOr}
\end{equation}
where $\mathcal{Z}=\sum_{n=1}^6 P_n$.

\subsubsection*{Model-Discussion}
To understanding how topological disorder affects the ability for particles to form a swarm we couple the disorder directly to the dynamics of the particles local information. This is achieved by allowing particles to only communicate with their n.n when a bond exists between them, determined by the conditional $\eta_{i}$. Treating disorder this way, missing bonds are represented as infinite barriers that forbid local information from influencing particle motion as well as inhibiting motion along that bond.  When $1-p>0$ the available amount of local information is limited as there is an average coordination number ($\bar{z}=pz$ , $z=6$ for a triangular lattice) available to each particle. In Fig.~\ref{fig: figure1}A we see an example of how a missing bond affects the next possible move that a particle can make. Consider the center particle in Fig.~\ref{fig: figure1}A and for the moment we will set $\epsilon$ to zero. If all lattice bonds were present (i.e all bonds are black) the particle would determine that the most probable direction would be to continue on along its current direction given by the grey horizontal arrow (along the $\mathbf{u}_4$ direction). In the case when the one bond is missing (grey) the particle has limited information about the local alignment field and will most likely choose to move along the white arrow, which is along the $\mathbf{u}_3$ direction. Furthermore, particles can passively run into a dead end where a bond is missing from the lattice. Once this happens motion ceases for that particular particle until a new direction is computed using Eqn.~\ref{eq:probOr} that takes the particle away from the broken bonds. For this reason broken bond directions are not allowed to be chosen by the Monte Carlo to avoid particles becoming permanently stuck at a broken bond. 

\subsection*{Finite Size Scaling}
To provide estimates for critical exponents we measured the system order parameter, $\langle v\rangle$ and the fluctuations of the order parameter, $\sigma^2_v$, for various system sizes at a fixed density, $\rho=1$. Using the finite size scaling functions below we extracted both the critical disorder fractions threshold and critical exponents (Fig. \ref{fig: figure2}B). To locate the critical disorder fraction in Fig.~\ref{fig: figure2}B we use the fact that the correlation length will scale as the system size near the critical disorder fraction, $\xi~L$, 
\begin{equation}
p^\ast(L) - p^\ast(\infty) \sim L^{-\lambda}
\end{equation}
where $\lambda=1/\nu$.

The susceptibility is related to the fluctuations of the order parameter near the critical disorder fraction,
\begin{equation}
\chi_v = \frac{N}{k_BT}\big[\langle v^2 \rangle - \langle v \rangle^2\big] =L^2\sigma^2_v.
\end{equation}
The peak of the susceptibility also scales as a function of system size (Fig. ~\ref{fig: supfig1}A) ,
\begin{equation}
\chi_v(p^\ast(L),L) \sim L^{\gamma/\nu}.
\end{equation}
To check the universality of these exponents over the range of repulsion magnitudes we fit both the order parameter and the susceptibility near the critical disorder fraction, $p^\ast(L)$(Fig. ~\ref{fig: supfig1}B and C).
\begin{equation}
\langle v \rangle \sim (p-p^\ast(L))^\beta
\end{equation}
and
\begin{equation}
\chi_v \sim (p-p^\ast(L))^{-\gamma}\sim L^2\sigma^2_v.
\end{equation}
Using these estimates we used the hyper scaling relation,
\begin{equation}
2\beta +\gamma=d\nu
\end{equation}
to compute $\nu$ when varying the repulsive interaction(Fig. ~\ref{fig: supfig1}D).

}

\setcounter{figure}{0} 
\makeatletter 
\renewcommand{\thefigure}{S\@arabic\c@figure}  
\makeatother

\begin{figure}[h]
\center \includegraphics[clip=true, trim=15 0 0 0,scale=0.5]{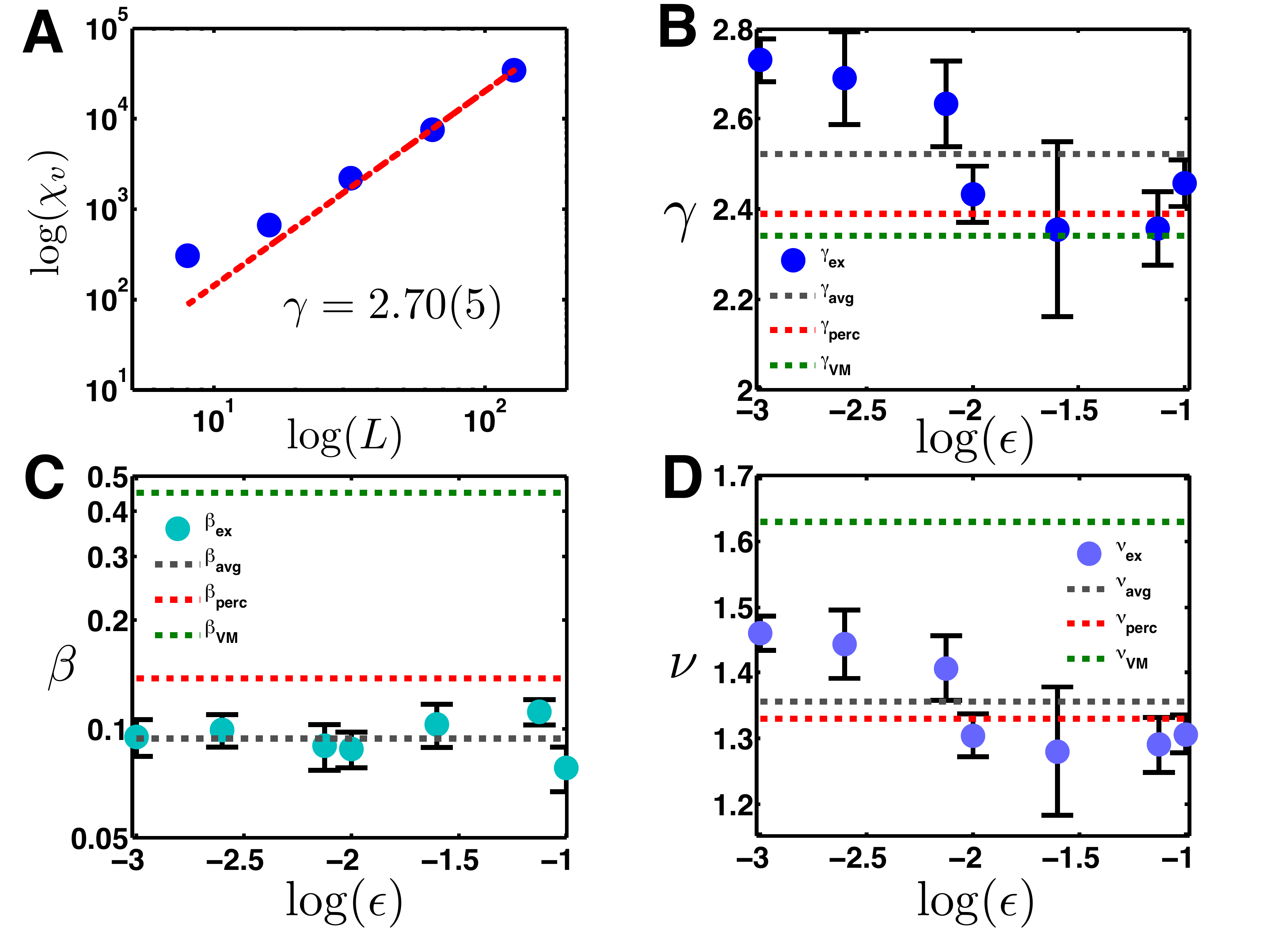} 
\caption{ (\textbf{A}) Finite size scaling fit (red dashed line)  of the susceptibility (blue filled circles) at finite repulsion ($\epsilon=10^{-1}$) as a function of system size. The critical exponent, $\gamma$, associated with the scaling of the susceptibility is shown. (\textbf{B}) Extracted critical exponents with error bars for $\gamma$ (blue circles and black bars) from scaling analysis near the critical disorder fraction transition ($1-p^\ast(L)$) plotted against the log of the repulsion magnitude. The grey dashed line is the average of the these values while the red line is the predicted exponent for percolation and the green line is the predicted exponent for the standard Vicsek model. (\textbf{C}) Filled circles are the extracted exponents, $\beta$ for the scaling of the order parameter near the critical disorder fraction [see supplementary material]. The dashed lines have the same color arrangement as the previous figure in the series. (\textbf{D}) Filled circles are the estimates for the critical exponent, $\nu$, associated with the correlation length, plotted over the magnitude of the repulsive interaction [see supplementary material]. Dashed lines indicate the averaged exponent and expected exponents from percolation and the standard Vicsek model. The color arrangement is the same as in the previous figures in this series.} 
\label{fig: supfig1}
\end{figure}

\begin{figure}[h]
\center \includegraphics[clip=true, trim=0 0 10 0,scale=0.5]{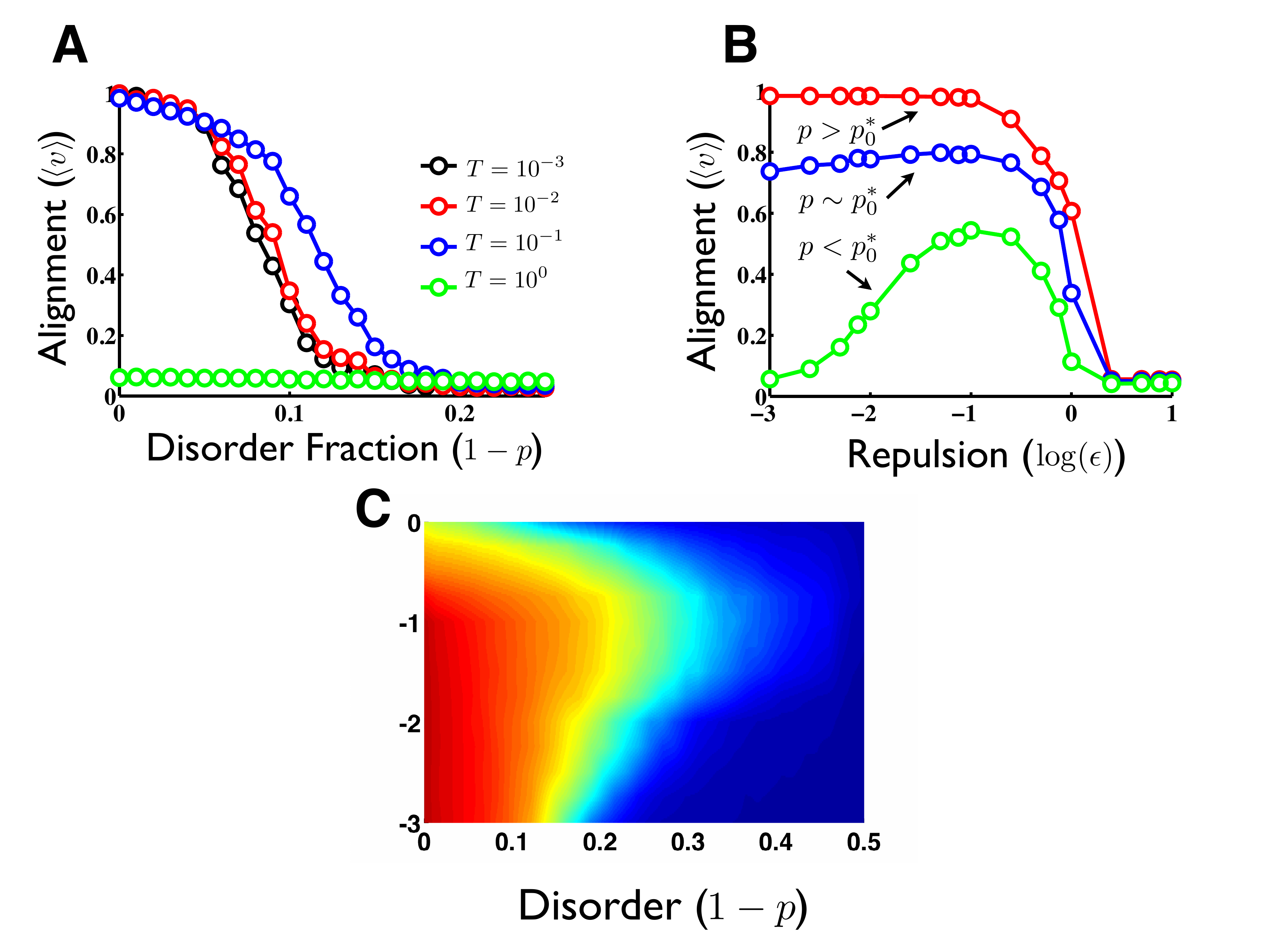} 
\caption{(\textbf{A})  Increasing thermal noise, for system without repulsion, but below the critical $T_c$ value there is an annealing effect that pushes the order-disorder transition to slight higher values of disorder fraction. (\textbf{B})The combination of both the repulsive energy and the temperatures near $T_c$ ($T=10^{-1}$) both help the ability for swarming behavior near the critical bond occupation probability $p_0^\ast$ for $\epsilon/\alpha \ll 10^{-1}$ as compared to Fig.~\ref{fig: figure3}A. (\textbf{C}) A phase diagram for larger thermal noise magnitude ($T=10^{-1}$) near the critical value $T_c$ (compare to Fig.~\ref{fig: figure1}C). The prominent peak near $\log(\epsilon)\sim -1$ has diminished significantly, but there also is a slight thermal noise effect which restores swarming for weak values of the repulsion magnitude ($\log(\epsilon) \sim -3$) and small values of the disorder fraction.}
\label{fig: supfig2}
\end{figure}

\begin{figure}[h]
\center \includegraphics[clip=true, trim=10 20 10 90,scale=0.8]{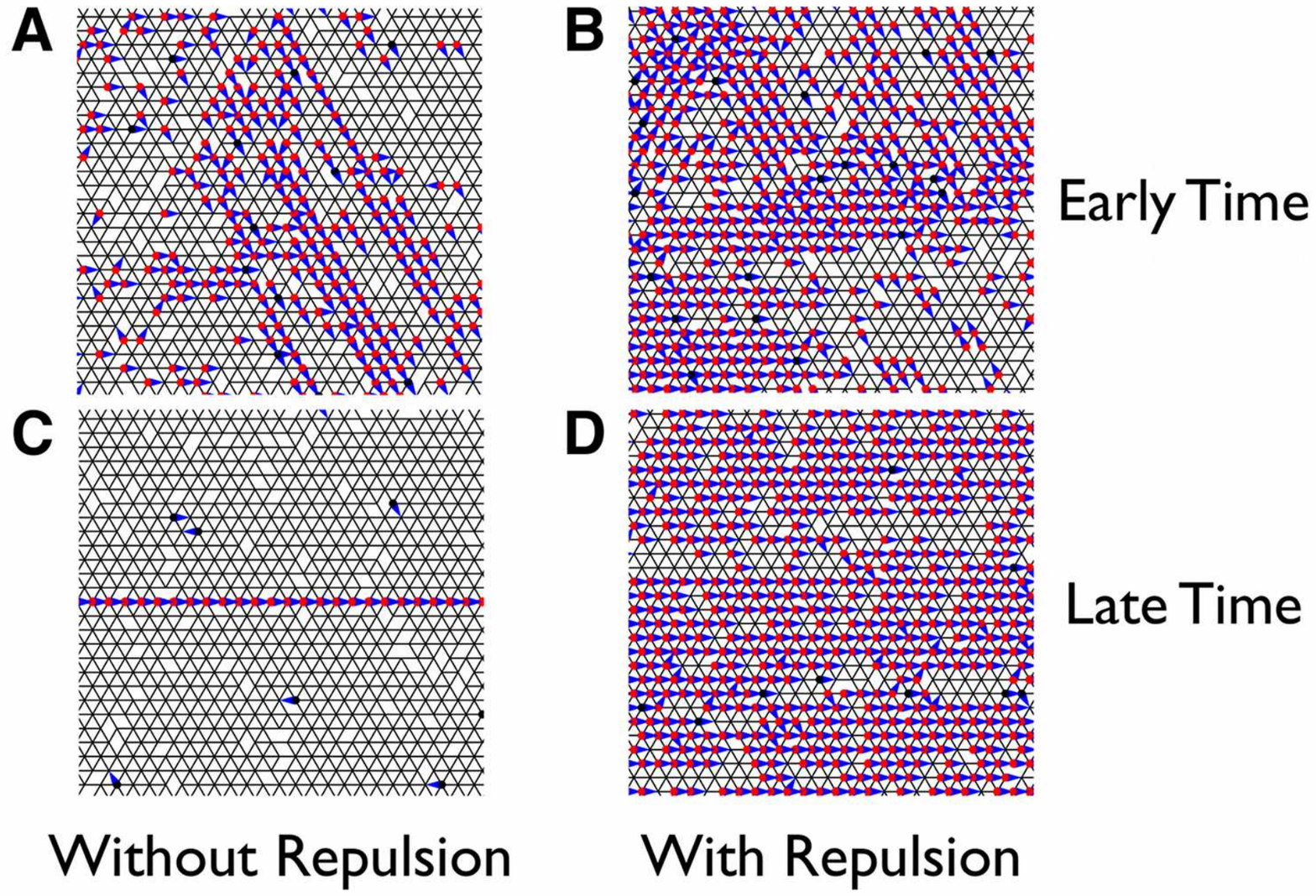} 
\caption{Snapshots of a simulations with and without repulsive energy.  (\textbf{A}) In early simulation times without repulsion, particles scatter off locations where bonds are missing and begin to form compact single file lines.   (\textbf{B}) With repulsion ($\epsilon=10^{-1}$) in early times, particles form extended groups that move together around defect bonds and fill in gaps caused by defects. (\textbf{C}) At late times without repulsion, particles form an extended single line that finds the a defect free lattice direction. Black particles are stuck at a broken bond indefinitely in this case. (\textbf{D})  In late times with repulsion, we find an collectively moving ordered swarm which can avoid defects and remain ordered by moving around particles which are temporarily stuck at a broken bond. All snapshots are for disorder fractions of $1-p=0.05$ at a thermal noise of $T=10^{-2}$}
\label{fig: supfig3}
\end{figure}

\bibliography{ref}
\bibliographystyle{apsrev}

\end{document}